# Can We Talk? An Exploratory Study of Gender and Network Ties in a Local Government Setting

Leisha DeHart-Davis
School of Government
The University of North Carolina at Chapel Hill
400 South Road
Chapel Hill, NC 27514
ldehart@sog.unc.edu

Nicole Humphrey
Department of Political Science
University of Miami
1300 Campo Sano Ave., 110E
Coral Gables, FL 33146
*Corresponding Author,* nhumphrey@miami.edu

Travis A. Whetsell
School of Public Policy
Georgia Institute of Technology
685 Cherry Street
Atlanta, GA 30332
travis.whetsell@gatech.edu

**Abstract:** We explore the influence of gender and formal organizational status on the formation of discussion ties. Network data, gathered through surveying employees from a municipal organization in the United States, garnered a 92% response rate (n=143). Results of exponential random graph modeling indicate women supervisors are more likely to send discussion ties, while women in general are more likely to receive discussion ties. These exploratory results suggest women may be perceived as more approachable for work discussions, but not as supervisors. Finally, the results identified a consistent homophily effect of gender in the discussion network.

**Acknowledgements:** Thank you to Meredith Newman for comments on an earlier draft.

**INTRODUCTION**

Researchers who study public sector organizations tend to focus on formal structure (Rainey et al., 2021; Whetsell, Kroll, & DeHart-Davis, 2020). While formal structure exerts a profound influence on its operations, public organizations also contain informal social networks that play an important role in the delivery of public goods and services (Kim & Lee, 2006). These social networks comprise constellations of employees who interact with each other, establishing informal "ties." Ties between employees—intraorganizational ties—provide information needed for work and policy implementation, while generating shared understandings of the work to help overcome complexity and uncertainty (Siciliano, 2015). Ties enable knowledge creation and dissemination, as well as organizational learning (Oparaocha, 2016; Paruchuri & Awate, 2017; Tasselli, 2015; Siciliano, 2017). This learning can be related to internal considerations (organizational processes) or external ones (community or political dynamics), both critical for public organizations in the delivery of public goods and services for addressing complex problems (e.g., COVID-19 and racial equity).

The communication and information sharing that are part and parcel of organizational learning are essential to effective human resource management performance at both the individual and organizational level, potentially making "the difference between government success and government failure" (Garnett, 1992, p. 3). To illustrate, social networks affect a range of human resource management activities, such as organizational change initiatives (Parise 2007), the transfer and storage of knowledge (Hollenbeck and Jamieson 2015), and organizational development (Hatala 2006). Thus, intraorganizational networks are another piece in the puzzle of effective human capital management rendered more poignant as performance expectations rise, budgets stagnate, and social problems remain wicked (Head & Alford, 2013). Despite the importance of intraorganizational social networks to effective human resource management, the extant literature in public administration has focused overwhelmingly on networks at the organizational level (Siciliano et al., 2021). Inter-organizational networks focus on the structures and processes of interaction between organizations, ignoring the internal dynamics of social interaction within organizations. This article builds on a growing body of intraorganizational networks literature in public administration, focusing on the internal communication network between employees of a small city government.

The importance of intraorganizational networks to the public sector raises the question of the attributes of organizational members that both send and receive ties. What types of individuals tend to reach out to others for information, and what types of individuals agree to provide that information? This is an important question from both scholarly and practical perspectives. From a scholarly perspective, much of the literature in public administration has focused on the structure and outcomes of inter-organizational networks (See Kapucu, Hu, & Khosa, 2017, for a review), with less attention paid to intraorganizational public sector network dynamics, much less the individual attributes of tie senders and tie receivers (Tasselli et al., 2015). From a practical human resources perspective, if internal communication is limited by social processes that bias on individual attributes (shared gender, race, age, etc.) it suggests that social networks are not as dense as they could be, thus impeding knowledge flow throughout the organization (Poleacovschi et al., 2021). Knowledge is central to developing and managing human capital within the public sector (Carnevale, 1996), whether through succession planning (Parisi 2007), professional development (Hatala 2006), or onboarding (Hollenbeck and Jamieson 2015).

This article focuses on one specific individual attribute in intraorganizational tie formation, the gender of tie seekers and tie receivers. Gender is a major organizing principle in organizational life (Acker, 1990) and thus is a logical individual attribute on which to base network tie formation. That gender could be a factor in network tie formation is suggested by four decades of research and theory about the role of gender in the formation of personal networks. In a sample of this evidence, women have been shown to have fewer friendship ties in organizations (Lincoln & Miller, 1979); to be located outside men's networks and perceived as less influential (Brass 1985); to receive less informal career help than men in networks (McGuire, 2002); to exhibit network behaviors that are somewhat different from men (Gremmen et al., 2013); and to be more likely to seek out men for advice and women for social support (Ibarra, 1992). These findings support the notion that women in the private sector have different and sometimes lower access to internal networks for enriching work and advancing careers. Our interest, by contrast, is in exploring the role of gender on tie formation as a public sector organizational phenomenon with implications for knowledge creation and dissemination. Specifically, does the gender of an organizational member influence the likelihood that they seek out or be sought out for network ties?

While there are several different types of network ties, e.g., friendship, advice, and trust (Tasselli et al., 2015), our study focuses on discussion ties in the workplace setting. Discussion ties occur when a person seeks out another person to discuss matters of importance to them (Marsden, 1987). Examining discussion tie formation entails looking at both incoming and outgoing discussion ties. Incoming discussion ties represent an employee being sought out to discuss work-related issues, while outgoing discussion ties represent an employee seeking out others to discuss work-related issues. Discussion ties fall into a broader category of network relationships called instrumental ties. In workplace settings, "instrumental ties arise in the course of work role performance and involve the exchange of job-related resources, including information, expertise, professional advice, political access, and material resources (Ibarra, 1993,

p. 59). In comparison, expressive ties are relationships grounded in friendship and social support, leading these contacts to be associated with higher levels of trust when compared to instrumental ties (Ibarra, 1993; Ibarra & Andrews, 1993). Prior research on gender and network formation has focused on exploring expressive ties (Feeney & Bernal, 2010; Quardokus & Henderson, 2015). We build on this literature and expand research on gender and network formation by studying an instrumental tie, which figures into public sector human resource management via mentoring (Parisi 2007), work unit cohesion (Hollenbrook and Jamieson 2015) and employee performance (Hatala 2006).

We draw on a range of theories to argue that gendered patterns of discussion ties will emerge in three ways within public organizations. First, women will be less likely to receive incoming discussion ties than men. Second, men will be less likely to initiate outgoing discussion ties than women. Third, supervisor status will moderate the relationship between gender and the formation of informal discussion-based ties. Specifically, we expect for women in supervisory roles to be sought out less for discussion ties, but also seek out discussion ties more compared to other organizational members.

We test these expectations in an exploratory study through a social network analysis of a small municipal organization in the Southeastern United States. The network data was collected using a survey of 143 employees, with a 92 percent response rate. Using exponential random graph modeling, we test the influence of gender and its interaction with supervisory status on the likelihood of both sending and receiving discussion ties. We control for race, age, supervisory status, trust, and several structural network processes. We do not evaluate the effect of intersectional identity due to low variability on race in the sample.

Through this exploratory analysis, our study makes several contributions to human resource management in the public sector. First, we explore gendered intraorganizational tie formation in the context of the public sector, which to the best of our knowledge has not been undertaken in the public management literature. Second, our use of network data and exponential random graph modeling allows for a more comprehensive view of employee interactions in the observed network. Finally, our approach is a novel contribution in that it melds two bodies of research, micro foundational network behavior (Tasselli et al., 2015) and workplace gender dynamics (DeHart-Davis et al., 2018). By connecting these areas of research, we help shed light on how gender shapes informal organizational networks, which is critical for understanding information sharing and performance within a public sector human resource management context.

**GENDER, SUPERVISORY STATUS, AND DISCUSSION TIES**

Network theory has often relied on a structural understanding of organizations when examining tie formation and network composition (McGuire, 2002). From this perspective, structural factors (e.g., formal organization status or supervisory status) are an important predictor of network development. Borrowing from Taselli and colleagues (2015), we take the alternative approach that information on the individual attributes of employees are also needed to understand how networks form. Research in this vein has sought to predict network formation as a function of individual attributes such as race and ethnicity (McPherson, Smith-Lovin, and Cook, 2001), personality (Landis, 2016), gender (Merluzzi, 2017), and education (Siciliano, 2015). A growing body of literature in public administration has examined the effects of individual attribute based homophily on workplace outcomes, e.g. effects of racial homophily on the perception of workplace inclusion (Jung and Welch, 2022) and race, gender, and occupational homophily effects on promotion (Marvel, 2021).

Building on previous literature, we explore the possibility that gender and its interaction with supervisory status have implications for incoming and outgoing tie formation. Extant research has shown that gender is central to how organizational members evaluate their coworkers and determine who should be in their network (Brass, 1985; Ibarra, 1997; McGuire, 2002; McDonald, 2011). Gendered stereotypes have the potential to influence which employees are perceived as worthy of receiving a tie (McGuire, 2000; 2002), as well as someone's comfort-level with seeking a tie with another organization member (Rosette et al., 2015). In the following sections, we provide more theoretical analysis of how incoming and outgoing discussion ties might be influenced by gender.

**Gender and Incoming Discussion Ties**

Incoming discussion ties depict an individual being sought out to talk through work-related issues of importance to the seeker (Marsden, 1987; Loscocco et al., 2009). When forming discussion ties, individuals seek out employees they perceive as competent and skilled, making expertise a notable catalyst of tie formation (Borgatti & Cross, 2003). This suggests that when compared to other employees, individuals sought out for discussion ties by their colleagues are considered valuable in terms of the information they yield (Nebus, 2006). We offer two theoretical reasons that women may be considered less organizationally valuable and thus less likely targets for discussion ties: the gendered constructions of expertise and the ascription of higher societal status to men.

While it is tempting to view expertise as an objective construct, indicated (for example) by advanced educational degrees or longer organizational tenure, feminist discourse indicates that expertise is a concept often framed by gender. Specifically, images of expertise rely on cultural masculinity, in that the expert is objective, autonomous, hierarchically a supervisor, and attached to a brotherhood of fellow experts (Stivers, 2002). Because these characteristics are culturally masculine and outside the realm of cultural femininity, women may be less likely to be perceived as experts, making them theoretically less desirable candidates for tie formation.

Another reason women may be less desired as discussion ties is that they are organizationally undervalued for a variety of reasons. Historically, public organizations divided masculine and feminine work (Guy & Newman, 2004) in ways that reflect the assumptions of agentic and communal attributes. Agentic attributes, commonly associated with men, emphasize qualities like independence, dominance, and rationality (Eagly & Johnson, 1990). Opposite of agentic attributes are communal attributes, which are often associated with women. Communal attributes emphasize characteristics like compassion and sympathy (Eagly & Johannesen-Schmidt, 2001). In sum, throughout organizations we should expect to find that gendered norms create stereotypes of how men and women should act (Mastracci & Bowman, 2015). Several scholars have suggested that these behavioral expectations have encouraged organizations to become gendered in a manner that prioritizes masculinity and places greater value on agentic behavior (Acker, 1990; Britton, 2000; Ely & Myerson, 2000; Stivers, 2002; Doan & Portillo, 2019).

The tendency of organizations to promote agentic attributes has consequences for how employees judge the value of their colleagues (Acker, 1990). Status characteristics theory suggests that employees evaluate the worth or competence of a colleague based on the social group membership of that colleague (Ridgeway, 2014; McGuire, 2002). Women potentially face being undervalued since they are often associated with communal attributes that do not align with historically masculine organizational priorities. In the context of discussion tie formation, this means that women may be seen as less desirable ties than men.

Scholars have conducted several studies providing empirical support for this assumption. Prior research has found that men have more homophilous networks than women, meaning their networks constitute like others, in this case, other men (Ibarra, 1992, 1997; Mehra, Kilduff, & Brass, 1998). Additional research suggests women are usually less central – with fewer connections to others -- in informal networks compared to men (Mehra et al., 1998; Ibarra, 1990; Lincoln & Miller, 1979; Miller, Labowitz, & Fry, 1975). Even when women are central in informal networks, they still struggle to be perceived as influential (Brass, 1985). In sum, women are often considered unideal targets for tie formation, even when they hold positions significant to the completion of organizational work. Following extant scholarship, we hypothesize that:

> *$H_1$: Women are less likely to be the recipient of incoming discussion ties than men.*

While theory and evidence suggest that women are less likely to receive discussion ties, it is important to note that some of the gendered theories contribute to an alternative rival hypothesis. Culture norms associate masculinity with attributes of independence, dominance, and rationality (Eagly & Johnson, 1990), while construing femininity in terms of being compassionate and sympathetic (Eagly & Johannesen-Schmidt, 2001), as well as nurturing and responsive (Eagly, Makhijani, & Klonsky, 1992). From emotional labor scholarship, women are often viewed as confidantes and caregivers (Martin, 1999; Guy & Newman, 2004). Research also suggests that women are construed as being more ethical, and by implication, trustworthy (Pandey et al., 2021).

While research has yet to examine gender differences in the perceived value of an individual person's time, women in supporting public sector roles have been required to operate inefficiently to accommodate the unique requests of male co-workers, one indication that their time is not valued equally (DeHart-Davis, 2009). And women in some support settings are perceived as more approachable (Bonnett & Alexander, 2012). This evidence supports the counterargument that women (Hochschild, 1983) could be *more* desirable targets for discussion ties.

### Gender and Outgoing Discussion Ties

In organizational settings, outgoing discussion ties represent an individual seeking out other employees to talk through work issues of importance. While discussion seeking can provide resources that are inaccessible when working alone (Bamberger, 2009), those resources come with a potential cost—a devaluation of one's competence by other organizational members (DePaulo & Fisher, 1980; Rosette, Mueller, & Lebel, 2015). Seeking help from a colleague suggests a shortcoming in one's understanding of a situation or work task (Borgatti & Cross, 2003). The individual seeking help potentially risks embarrassment for failing to be as knowledgeable as other employees (Nebus, 2006). Due to gendered social expectations, previous research has found the organizational costs of seeking out information can be more detrimental for men than women (Lee, 1997; 2002).

Exploring the alignment between gender and professional roles, role congruity theory suggests that masculinity emphasizes dominance, assertiveness, and independence (Rosette et al., 2015; Collinson, 2010). When an individual seeks out others for discussion, they acknowledge their own lack of understanding and reliance on the employee providing them with information (Lee, 1997), behavior not aligned with expectations of masculinity. In line with this argument, a study of engineers showed that men perceived higher social costs of seeking out women for discussion compared with their male counterparts (Poleacovschi et al., 2021). In another strand of reasoning, men have been shown to be overconfident in their knowledge in a number of realms, including financial investments (Barber & Odean, 2001; Mishra & Metilida, 2015) and classroom knowledge (Lundeberg et al., 1994; Bentsoon et al., 2005). Thus, gender differences in outgoing discussion ties may be driven, in part, by men's overconfidence in their knowledge of workplace situations. Potential threats to masculinity for seeking out help and the masculine tendency to be overconfident in one's knowledge suggests men will be less likely to seek others out as discussion ties compared to women (Rosette et al., 2015).

Organizational research suggests women are generally more likely to seek help than men. Lee (1997) hypothesized that gender socialization would lead women to seek help more than men based on an "other" orientation that encouraged close relationships and shorter interpersonal power distances. Bamberger (2009) observed a range of research that explains the greater likelihood of women to seek help because women are not socialized to privilege power or perceived competence. Men will place a greater emphasis on the potential negative perception that help-seeking creates compared to women (Bamberger, 2009). Women have also been hypothesized to seek help more so than men due to their lower confidence levels (an expectation not born out by experimental evidence, Heikensten & Isakson, 2019). Another theoretical explanation for women seeking others for discussion is that doing so exhibits an interdependence and communality that is culturally feminine (Poleacovschi et al., 2021). Gender stereotype theory has been used to explore perceptions that women and men give and prefer receiving different kinds of help, driven theoretically by stereotypes of women as helpless and men as competent (Chernyak-Hai & Waismel-Manor, 2019).

Overall, theory and empirical research suggests that gender will influence an individual's willingness to initiate a discussion tie in the workplace. Because prior research suggests that men experience gender role expectations that render self-reliance a culturally masculine expectation (Rosette et al., 2015) and women are socialized to close power distances and not privilege perceived competence (Lee, 1997; Bamberger, 2009), we hypothesize the following:

*H$_2$: Men will be less likely than women to initiate outgoing discussion ties.*

**Gender, Supervisor Status, and the Formation of Discussion Ties**

Possessing supervisory status (i.e., holding a supervisor position) brings a credibility, authority, and status that should influence network behavior (Kanter, 1975; Brass, 1985; Luo & Cheng, 2015). From the perspective of incoming tie formation, individuals with supervisory status should be sought out more because they have greater resources to offer (McGuire, 2002). Conversely, supervisory status should reduce the need to reach out to others to discuss work-related issues because that credibility and authority should provide higher-level information that precludes searching for information (Liu & Moskvina, 2015).

But formal organizational status is not a gender-neutral proposition because being a woman complicates power dynamics. Beginning with incoming discussion ties, powerful women are likely to be socially penalized as leaders because holding power violates standards of feminine communal behavior—nurturing, compassionate and responsive (Eagly, 1993; Johnson et al., 2008)—and instead conveys agentic behaviors (Eagly et al., 1992; Williams & Tiedens, 2015). Women incur gender penalties by merely assuming a leadership role, as men in leadership roles tend to be evaluated more favorably than women. This advantage is intensified in environments dominated by men (Eagly et al., 1992; Funk, 2019). There is growing evidence that women are perceived as less likeable (Rudman, 1998; Williams & Tiedens, 2015) and more uncooperative by colleagues (Bowles, Babcock, & Lai, 2007) when presenting agentic behavior. For female supervisors, the negative effects of breaking with gendered stereotypes by serving in a leadership role may lead to being perceived as less desirable for incoming discussion ties. In support of this argument, a study of engineers suggest that men perceive higher social costs – including greater vulnerability, nervousness, and discomfort – when seeking knowledge from women (Poleacovschi et al., 2021). In short, we anticipate women with supervisory positions maybe perceived negatively by others and elicit avoidance from organizational members, reducing their number of incoming discussion ties. This leads to our third hypothesis, that:

*$H_3$: Women in supervisory roles will be less likely than other organization members*

*to be sought out for discussion ties.*

When considering outgoing discussion ties for women in supervisory positions, gendered leadership theory provides insight into why powerful women are more likely to initiate discussion ties. Scholars have found that women as leaders tend to demonstrate democratic leadership behaviors more than men (Eagly & Johnson, 1990). This form of leadership has been found to be associated with higher participation among employees (AbouAssi & An, 2017) due to its emphasis on communication and inclusivity (Fox & Schuhmann, 1999; Hamidullah, Riccucci, & Pandey, 2015). A study of employees across several sectors found that both men and women felt that women managers in their organizations displayed higher levels of interpersonal and collaborative skills compared to men (Chesterman, Ross-Smith, & Peters, 2005). By contrast, men have been found to promote hierarchy and top-down communication with subordinates (Eagly & Johnson, 1990). This scholarship on gendered leadership leads our fourth hypothesis:

*H₄: Women in supervisory roles will be more likely than other organization members to seek out discussion ties.*

Collectively, these hypotheses highlight the potential significance of gender and supervisory status when exploring tie formation. Empirical support for these hypotheses will imply that the emergence of informal discussion networks within public sector organizations—and by extension, organizational learning, and knowledge flow—is shaped by gender, as well as the interaction of gender and the formal leadership status of organizational members.

**RESEARCH METHOD**

To estimate the effects of gender and supervisor status on the formation of discussion ties, we distributed a Qualtrics survey to the 155 employees of a small coastal local government in the Southeastern United States. This local government comprised ten departments of various sizes and functions. The participating local government volunteered to participate based on their desire to understand intra-organizational communication flows. We conducted a survey pre-test with a small group of employees and used that feedback to incorporate minor changes into the survey instrument.

To incentivize participation, the local government offered four hours vacation time to research participants randomly selected by the research team. The survey was administered in October 2019 and remained open for two weeks. Before taking the survey, all participants confirmed their informed consent of participation. The response rate was 92 percent (n=143/N=155). Using a sample with such a high response rate makes this case more ideal because we can have a higher level of confidence that the analysis reflects the actual network within the organization. Median survey scores were imputed for two individuals who only partially completed the survey. The survey sample represents the city workforce in age, gender, and race/ethnicity, and departmental representation. However, the sample was relatively homogenous, resulting in low variability on race with respect to supervisory status.

Social network data were generated using the roster method, which asks every employee to identify their network relationships with every other member from a complete list (Wasserman & Faust, 1994; Wald, 2014; Perry, Pescosolido, & Borgatti, 2018). Specifically, survey participants were asked to indicate whether they sought out each organizational member to "discuss work-related matters." The language "seek out" was used to establish directionality to the network ties. This item appeared alongside two other network items, which asked about whether the employees seek out other organizational members for information to do their jobs or permission regarding work tasks. Hence, the survey instrument differentiated discussion ties from information and permission seeking, which represent more formal connections. Organizational members were presented by department to make indicating interactions easier. An example of this question framing is provided in Appendix 1.

Our dependent variable is binary, the presence or absence of a discussion tie, measured as "1" if a tie is present or "0" if a tie is absent. Gender is measured using a dichotomous variable with 0 representing men and 1 representing women. While the data for this study includes only two categories to operationalize gender (i.e., man or woman), we recognize that gender is a socially constructed identity and exists along a spectrum that is not limited to two categories. Supervisory status is measured by a dichotomous variable indicating 0 for non-supervisory status and 1 for supervisory status. A multiplicative term between gender and supervisor is created to assess the interaction between gender and supervisor status.

Our demographic control variables include age and race. Trust was included to account for the effect of an employee's willingness to be vulnerable in discussion seeking; it is a common variable in social network analyses (as one example, see Shazi et al., 2015). Trust was measured using three survey items combined into a single principal component. The items were designed to get at the trust between employees and supervisors, e.g., "In my department, employees trust supervisors". (See Appendix 1.2 to view this survey question). Further, it is particularly important to control for organizational structure in intra-organizational networks (Whetsell, Kroll & DeHart-Davis, 2021) since exchange relationships can be very different within organizational sub-units. The sample included 10 departments across the organization with considerable variation in size and composition. Thus, we include three model terms capturing departmental structure, including departmental tie receiver effects (the extent of incoming ties to individuals within a given department), departmental tie sender effects (the extent of outgoing ties from individuals within a given department), and departmental homophily effects (the degree to which individuals seek out individuals within the same department). The departmental tie receiver and sender effects are included as fixed parameters in the models. We also included two edge covariates – other types of ties, e.g., permission and information seeking -- which reveal formal organizational relationships (Tasselli et al., 2015). We also account for several properties of network structure commonly used in social network studies, including density (the proportion of actual to possible relationships in a network), reciprocity (the extent to which ties are reciprocated between individuals), transitivity (friends of friends), and activity spread (the distribution of ties) (Robbins, 2007; Lusher, Koskinen & Robins, 2013). These structural terms are important to include as network controls, since for example gender or racial homophily may be overestimated in the absence of reciprocity or transitivity terms (Wimmer & Lewis, 2010).

Table 1 contains descriptive statistics for the sample. Women comprise 29 percent of the sample and people of color (POC) comprise 8 percent of the sample. The average participant is 42 years of age, with employees ranging from 18 to 77 years old. Thirty-two percent of the sample are made up of supervisors. Among the supervisors only 8 are women and 3 are POC.

**Table 1. Descriptive Statistics**

|  | Sum | Minimum | Maximum | Mean | Std. Deviation |
| --- | --- | --- | --- | --- | --- |
| Women | 42 | 0 | 1 | 0.29 | 0.46 |
| POC | 12 | 0 | 1 | 0.08 | 0.28 |
| Age | na | 18 | 77 | 42 | 12 |
| Supervisors | 46 | 0 | 1 | 0.32 | 0.47 |
| Women Supervisors | 8 | 0 | 1 | 0.06 | 0.23 |
| POC Supervisors | 3 | 0 | 1 | 0.02 | 0.14 |
| Trust | na | -3.04 | 1.27 | -0.00 | 1 |

To investigate the effects of gender and supervisor status on the formation of discussion ties and the emergence of a social network, we employ exponential random graph modeling (ERGM). ERGM is a technique developed to account for dependencies in observations that arise in relational data (Robins et al., 2007). Traditional techniques such as logistic regression assume independence of observations, in which each observation is like a coin toss where the odds of seeing heads or tails is independent of the previous coin toss (Wasserman & Pattison, 1996). The independence assumption is violated in the network context, where node values are *inter*dependent and tie formation between nodes is contingent on broader network processes. Thus, testing hypotheses regarding the drivers of tie formation is typically biased in the traditional regression setting.

ERGM by contrast accounts for interdependent observations in the data by modeling network processes such as popularity (an individual's number of ties), reciprocity (the extent to which ties are reciprocated), homophily (similar people forming ties with one another) or transitivity, individuals forming ties with each other based on a shared third individual (Goodreau, Kitts, & Morris, 2009). Further, ERGM is useful for estimating the probability of tie formation between individuals in an observed network based on attributes such as gender, race, etc., as well as estimating the effect of shared attributes between individuals such as homophily (alike in some respect) and heterophily (different in some respect). ERGM provides parameter estimates for variables of interest by creating a distribution from thousands of simulated networks based on the characteristics of the observed network (Hunter et al., 2008). Parameter estimates for ERGMs are interpreted similarly to log odds estimates in logistic regression models.

To estimate goodness-of-fit, the model undergoes an iteration process which converges when the observed network is probable given the distribution of simulated networks (Lusher et al., 2013). Recently, ERGMs have become more commonly used for analyzing social networks in public administration research (e.g., Siciliano, 2015; Nisar & Maroulis, 2017; Whetsell et al., 2021).

**RESULTS**

First, the results present the network visualization, as well as descriptive statistics on gender differences in the network, including network centrality, defined as how central an individual is in a network. The intra-organizational discussion network contains 143 individuals and 3815 directed discussion ties. The network is quite dense with a network density of 0.187. Figure 1 shows a visualization of the discussion network, color coded as men (blue)/women (red), and supervisor (darker)/non-supervisor (lighter) status. The size of the nodes reflects the in-degree of the node, i.e., the number of ties directed at the individual. The layout of the nodes and edges uses the Fruchterman-Reingold algorithm which is a force-directed layout based on the principle of gravity, where more well-connected individuals are placed further toward the center of the network and less well-connected individuals appear on the periphery (Fruchterman & Reingold, 1991). Upon visual inspection, there does not appear to be significant centralization (in which certain individuals control discussion flows) or clustering (groups of people who form discussion clusters distinct from the rest of the network).

**Figure 1. Network Visualization City Government Discussion Network**

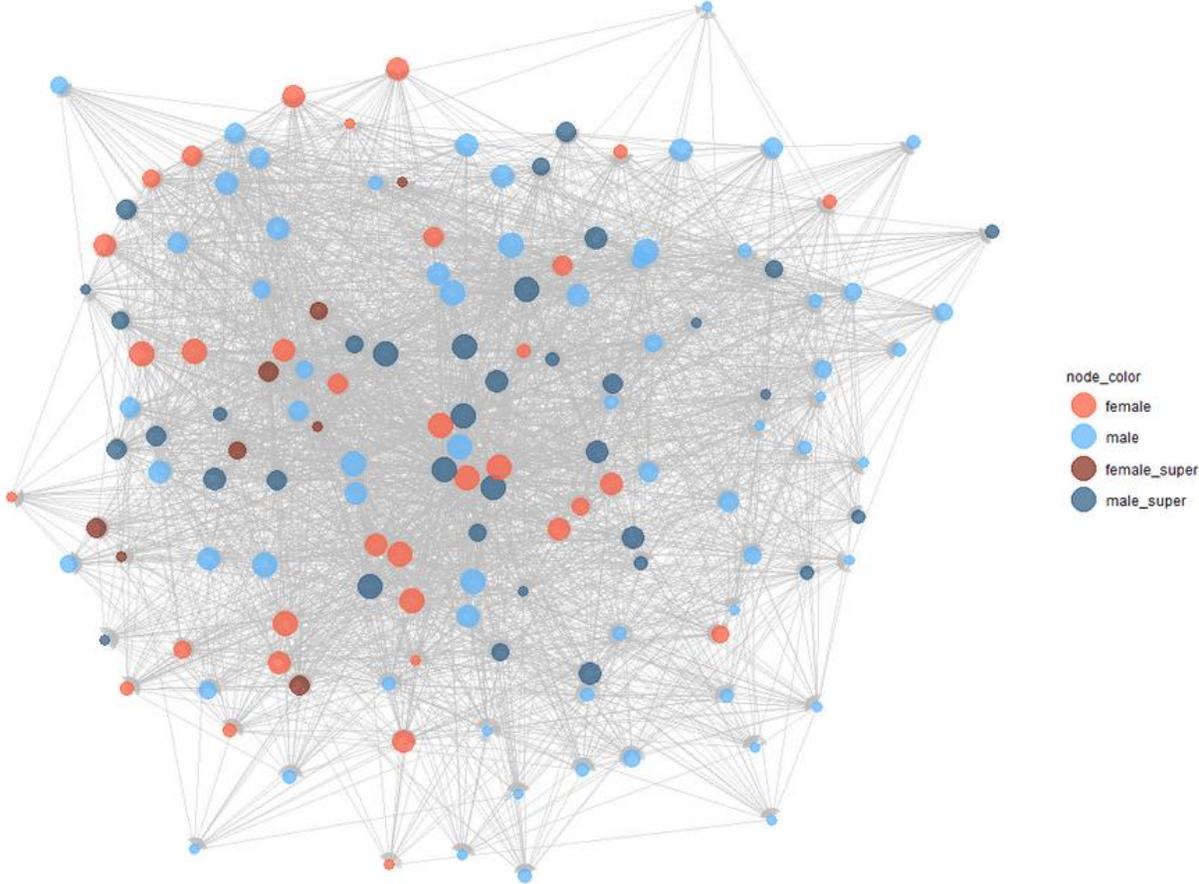

There are 48 women in the organization with 8 of them being supervisors, while men constitute 97 members of the organization, 34 of which are supervisors. This distribution is not unusual in city government, which tends to be male dominated by virtue of departments such as police, fire, and public works (Guy, 2017). Hence, the organization itself is disproportionately staffed by men, who are also more likely to be supervisors. Before moving to the ERGM analysis, we begin with a basic descriptive analysis comparing network characteristics based on gender and supervisory status. The Wilcoxon test for non-parametric data is used to generate p-values since the distributions of network centrality measures tend to non-normal. Table 2 shows the mean comparisons for in-degree centrality, out-degree centrality, eigenvector centrality, or betweenness centrality, four measures common statistics used to characterize network centrality (Wasserman & Faust, 1994). In-degree measures the number of ties received by the individual, while out-degree measures the number of ties sent by an individual. Eigenvector centrality measures the degree to which an individual is connected to many other well-connected individuals (Bonacich, 1987). Betweeness measures the degree to which an individual falls on a path between other individuals in the network (Freeman, 1977). We find no general statistically significant difference in network centrality between men and women on in-degree centrality, out-degree centrality, eigenvector centrality, or betweenness centrality. However, we do find significant differences regarding the interaction between supervisory status and gender identification. More specifically we find that women supervisors have higher out-degree centrality and higher eigenvector centrality than everyone else in the organization. Similarly, we find that men have higher out-degree, eigenvector, and betweenness centrality than everyone else. However, when matched against each other (omitting the rest of the staff from the sample) there are no significant differences revealed by the analysis. This may be due partly to the much smaller sample size after omitting non-supervisors.

**Table 2. Comparison of Network Centrality by Women/Men and Supervisory Status**

|  | In-Degree | Out-Degree | E.Centrality | B.Centrality |
|---|---|---|---|---|
| Men | 25.535 | 26.772 | 0.069 | 131.725 |
| Women | 28.786 | 25.810 | 0.068 | 124.185 |
| Wilcoxon Test P-Value | 0.136 | 0.835 | 0.736 | 0.828 |
| Everyone Else | 26.837 | 25.830 | 0.067 | 129.110 |
| Women Supervisors | 20.625 | 37.625 | 0.100 | 136.266 |
| Wilcoxon Test P-Value | 0.280 | 0.027 | 0.029 | 0.480 |
| Everyone Else | 26.190 | 23.410 | 0.061 | 104.782 |
| Men Supervisors | 27.316 | 35.000 | 0.088 | 197.840 |
| Wilcoxon Test P-Value | 0.554 | 0.005 | 0.009 | 0.003 |
| Men Supervisors | 27.316 | 35.000 | 0.088 | 197.840 |
| Women Supervisors | 20.625 | 37.625 | 0.100 | 136.266 |
| Wilcoxon Test P-Value | 0.207 | 0.354 | 0.310 | 0.619 |

Table Notes: In-degree measures the number of incoming ties or ties received by the individual, while out-degree measures the number outgoing or ties sent by an individual. Eigenvector centrality measures the degree to which an individual is connected to many other well-connected individuals. Betweeness measures the degree to which an individual falls on a path between other individuals in the network.

Next, we present the results of the exponential random graph models in Table 3. The table presents three models. Each variable of interest includes a receiver effect, a sender effect, and a homophily effect, the latter of which estimates the likelihood that similar types of individuals will interact through discussion ties. All three terms are included for each variable, which is common practice in ERGM analysis for directed networks. The first model includes the demographic variables for gender, age, and race, departmental effects, permission, and information seeking behavior, and includes important structural network control variables that account for social processes of network formation. The second model adds the supervisory status variable. The third model adds an interaction term between gender and supervisory status. All estimates are interpreted as the log odds of a tie forming between two nodes, like a logistic regression model. Significant independent variables are indicated by parameter estimates that are at least twice the size of standard errors, a ratio reflected in the Wald test statistic (Lusher et al., 2013, p. 157).

**Table 3. Exponential Random Graphs Models – Gender and Supervisor Status**

|                              | Model 1             | Model 2             | Model 3             |
|------------------------------|---------------------|---------------------|---------------------|
| Woman(receiver)              | 0.116 (0.056)*      | 0.093 (0.057)       | 0.219 (0.063)***    |
| Woman (sender)               | 0.048 (0.029)       | 0.060 (0.031)       | 0.047 (0.034)       |
| Woman (match)                | 0.170 (0.046)***    | 0.172 (0.047)***    | 0.159 (0.048)***    |
| Supervisor(receiver)         |                     | -0.108 (0.049)*     | 0.017 (0.053)       |
| Supervisor(sender)           |                     | 0.088 (0.028)**     | 0.071 (0.031)*      |
| Supervisor(match)            |                     | 0.001 (0.044)       | -0.015 (0.044)      |
| Woman Supervisor(receiver)   |                     |                     | -0.252 (0.174)      |
| Woman Supervisor(sender)     |                     |                     | 0.417 (0.147)**     |
| Woman Supervisor(match)      |                     |                     | 0.392 (0.158)*      |
| Department(receiver)         | Fixed               | Fixed               | Fixed               |
| Department(sender)           | Fixed               | Fixed               | Fixed               |
| Department(match)            | 0.427 (0.059)***    | 0.433 (0.058)***    | 0.428 (0.058)***    |
| EdgeCov(permission network)  | 2.400 (0.086)***    | 2.413 (0.088)***    | 2.419 (0.091)***    |
| EdgeCov(information network) | 0.179 (0.061)**     | 0.174 (0.063)**     | 0.178 (0.064)**     |
| Age(receiver)                | 0.008 (0.002)***    | 0.009 (0.002)***    | 0.008 (0.002)***    |
| Age(sender)                  | 0.004 (0.001)***    | 0.003 (0.001)**     | 0.003 (0.001)**     |
| Age(difference)              | -0.006 (0.002)**    | -0.005 (0.002)**    | -0.005 (0.002)**    |
| POC(receiver)                | 0.490 (0.127)***    | 0.459 (0.131)***    | 0.564 (0.128)***    |
| POC(sender)                  | 0.290 (0.103)**     | 0.301 (0.105)**     | 0.303 (0.104)**     |
| POC (match)                  | 0.423 (0.124)***    | 0.414 (0.123)***    | 0.440 (0.121)***    |
| Trust(receiver)              | 0.065 (0.024)**     | 0.072 (0.024)**     | 0.056 (0.024)*      |
| Trust(sender)                | 0.030 (0.012)*      | 0.029 (0.014)*      | 0.027 (0.013)*      |
| Trust(difference)            | 0.015 (0.020)       | 0.016 (0.021)       | 0.016 (0.020)       |
| Edges                        | -1.344 (0.243)***   | -1.412 (0.254)***   | -1.926 (0.287)***   |
| Mutual                       | 0.005 (0.066)       | 0.021 (0.068)       | 0.024 (0.067)       |
| gwesp.OTP.fixed.0.5          | 0.233 (0.071)**     | 0.258 (0.073)***    | 0.264 (0.072)***    |
| gwodeg.fixed.3               | -3.338 (0.064)***   | -3.273 (0.069)***   | -3.283 (0.071)***   |
| AIC                          | 16647.765           | 16625.411           | 16604.478           |
| BIC                          | 16940.756           | 16942.158           | 16944.981           |
| Log Likelihood               | -8286.882           | -8272.706           | -8259.239           |

Table Notes: ***p < 0.001; **p < 0.01; *p < 0.05, standard errors in parentheses. POC refers to person of color. Receiver is the effect of the variable on the likelihood of receiving a tie. Sender is the effect of the variable on the likelihood of sending a tie. Match is the likelihood of two nodes forming a tie on the basis of a shared categorical attribute on the variable. Difference is the likelihood of a tie forming on the basis of the difference between values on a continuous variable. Edges models the density of the network. Mutual models reciprocity. Gwesp models transitivity. Gwodeg models popularity. A fourth model was conducted on BIPOC*Supervisors which had non-significant results on the estimates and left main effects unchanged.

Results of the first model fail to support our first two hypotheses. Women are *more* likely than men to receive ties. Furthermore, there is no statistically significant difference between women and men in the likelihood of initiating outgoing decision ties, thus failing to confirm our second hypothesis, that men will be less likely to initiate outgoing discussion ties. There is a significant homophily effect of gender in the network, i.e., women tend to form ties with women and men tend to form ties with men, which is consistent with the network literature. Homophily is also observed for employees of color and employees the same age.

The second model, which adds supervisors, slightly changes the base model. Gender differences in incoming discussion ties by organizational members are no longer significant and gender differences in outgoing discussion ties remain insignificant, thus failing to support our first and second hypotheses. The model results also suggest that supervisors are more likely to send discussion ties, but less likely to receive discussion ties than other organizational members. There is no homophily effect for supervisors.

The third model incorporates the interaction between gender and supervisor status. As in the first model, women are more likely than men to receive discussion ties, but there is no gender difference in the initiation of outgoing discussion ties. Thus, our first hypothesis—that women will be less likely to receive discussion ties—is contradicted and our second hypothesis (that men will be less likely to initiate discussion ties) is unsupported. Our third hypothesis, that women supervisors are more likely than other organizational members to send discussion ties, is supported. However, there is no statistically significant difference in incoming ties between women supervisors and other organizational members. Finally, there appears to be a significant homophily effect for women supervisors, while the gender homophily among organization members seen in the first two models persists in the third model.

**DISCUSSION**

This article sought to explore the influence of gender and supervisor status on the formation of discussion ties in a local government organization. Using data collected from a social network analysis of a small coastal town in the Southeastern United States, (n=143, 92 percent response rate), we model the formation of incoming and outgoing discussion ties as a function of gender and supervisory status, accounting for a range of controls that include age, race, and network structure variables. Discussion ties are an important network behavior, as they depict employees discussing matters of importance to them regarding their work. Discussion ties are also carriers of information with implications for a range of human resource management issues, including knowledge management, mentorship, workplace climate, and succession planning. Thus, our examination of the gendered nature of discussion tie formation is timely and relevant.

We first hypothesized that women within the organization would receive fewer incoming ties than men. This expectation was based on theories of the gendered nature of expertise (Stivers, 2002), as well as the relationship between gender and social status (Ridgeway, 2014), which assign less importance and value to the feminine role and theoretically make it more difficult to attract incoming ties. Our findings generally indicate the opposite, that women receive *more* discussion ties than men. One potential explanation for this contradictory finding is gendered assumptions that women are nurturing, responsive, and trustworthy confidantes (Hochschild, 1983; Martin, 1999; Guy & Newman, 2004) may lead other employees to feel more comfortable seeking out women for discussion. It is also possible that this higher likelihood of being the recipient of discussion ties represents emotional labor being disproportionately outsourced to women, along with the additional costs of time spent in the name of responsiveness. Should this result hold in future research, qualitative methods could be used to determine potential interim mechanisms leading to these tie formation patterns, be they culturally feminine characteristics, the nature of occupational roles held by women, or perceived trustworthiness.

Our second hypothesis argued that men would be less likely to initiate outgoing discussion ties. This hypothesis was based on the theoretical assumption that discussion seeking requires a willingness to be vulnerable in seeking out others. Such vulnerability, we argued, potentially threatens norms of cultural masculinity, such as competence and independence (Collinson, 2010), while possibly supporting the feminine norms of interdependence and lower power distance (Lee, 1997, Bamberger, 2009). Evidence of higher levels of overconfidence among men (Mishra & Metilida, 2015) contributed to our expectation that men would be less likely to seek out discussion ties. This finding could be a function of an organizational culture that emphasizes learning and thus makes it safe (even necessary) for men to seek help. Should future findings prove consistent, the result is positive from a practical perspective because discussion seeking can provide employees with resources and information (Bamberger, 2009) that can help them complete their work and advance their careers. If men in this organization were resistant to seeking out discussion ties, it could create deficiencies in organizational knowledge. Future research should seek to test this exploratory finding given the significance of knowledge creation and dissemination for effective public sector organizations.

Our third hypothesis is partly supported. The first part of the hypothesis contended that female supervisors would receive fewer incoming discussion ties compared with other organization members. This hypothesis was based on theory and evidence that women with organizational power convey agentic behavior (Eagly et al., 1992; Williams & Tiedens, 2015) and thus elicit pushback, such as not being sought out to discuss work-related issues. Our results indicate that, in our sample, female supervisors (nor supervisors in general) differed from other organization members in the probability of receiving incoming discussion ties. While this result contradicts our expectation, it may mean that employees are reluctant to seek out those with formal organizational status for informal discussions. If this finding holds consistent in future research, investigators should examine whether it reflects a lack of employee voice, where supervisors encourage employees to raise concerns or ideas for improvement (Morrison, 2014).

The second part of our third hypothesis, that female supervisors would be more likely to initiate discussion ties, is supported. We formulated this expectation based on theory and research that women on average employ more inclusive and consultative leadership styles than men (Fox & Schuhmann, 1999; Hamidullah et al., 2015). While it appears that women in supervisor positions reach out more than the average employee, it could be that supervisory women see the necessity in reaching out to obtain information that is not forthcoming through incoming discussion ties, which can yield valuable organizational intelligence. While this pattern is not unique to female supervisors—all supervisors tend to reach out more than other organizational members—it is more pronounced. Future research should examine if this pattern still holds and, if so, determine whether supervisors are reaching out more to their subordinates (potentially suggesting close supervision) or to their superiors (potentially suggesting centralized decision-making).

Collectively, these results may point to a higher level of informal interaction required of women in public organizations. At the employee level, women are disproportionately targeted for incoming ties; at the supervisory level, women disproportionately seek out others for discussion. On the upside, these patterns suggest that women play a unique role in informal social networks and the organizational learning they produce. The downsides may be lost opportunities for organizational learning if men are less likely to be sought out for discussion ties or if male supervisors are less likely to reach out to others to learn. While our results suggest that gender sometimes influences network tie formation on human resource management issues such as knowledge management, succession planning, mentoring, and organizational climate. In addition, scholars should explore how communal and democratic leadership styles versus transactional and hierarchical leadership styles influence network tie formation. The latter research queries will suggest the extent to which leaders undercut or foster learning and knowledge dissemination within organizations.

At a broader level, there has been a substantial increase of network-focused research in public management over the past several decades (Kapucu, Hu, & Khosha, 2017). Most of these studies have explored inter-organizational networks focused on policy and collaboration (Kapucu et al., 2017; Lecy et al., 2014, Isett et al., 2011). While public management scholars have begun exploring intra-organizational networks (See Siciliano 2015, 2017; Whetsell et al., 2021), research examining the experiences of individual employees lacks in comparison to group-level analyses. This study builds on a growing body of network research within the field public management to provide helpful insights on network formation *within* organizations.

This research also contributes broadly to studies of network formation at the micro-level (see Siciliano, Wang, and Medina 2020), which Tasselli and colleagues note can be understood using three perspectives (2015). First, the individual actor-oriented perspective is a psychological approach that suggests individual characteristics influence network formation (Kilduff & Krackhardt, 1994). Inversely, the second approach is a structural sociological perspective suggesting that individual actors are influenced, constrained, and enabled by the network (Burt, 1982; Wellman, 1988). Finally, the coevolutionary perspective is a combination of social and psychological approaches that interacts individual characteristics with network structure and processes, suggesting both actors and networks evolve simultaneously (Coleman, 2000; Kossinets & Watts, 2009). This study's findings suggest the importance of a co-evolutionary perspective by advancing a model that includes individual attributes with a primary focus on gender and network characteristics. While public management scholars have often relied on structural perspectives to explore inter-organizational networks, studies of informal intra-organizational networks that include attention to individual attributes are far less common in the literature (Isett et al., 2011; Kapucu et al., 2017). Our research shows that incorporating individual and structural characteristics provides additional nuance to the field's understanding of intra-organizational networks. In addition, our use of network frameworks grounded in psychology, sociology, organizational studies, and public management answers longstanding calls for "cross-fertilization" across research disciplines when conducting network analysis (Berry et al., 2004).

The primary strength of this study is a 92 percent survey response rate, which provides a suitable case to explore gender dimensions of discussion tie formation. The high-level of participation among employees means we can be more confident that the information gathered from this study reflects the actual network within the organization. However, while single-organization research is common in social network analysis because of the requirement for organizational access and labor-intensive data collection (Flap et al., 1998). And while the gender distribution of this small city mirrors local government demographics generally -- characterized by mostly white men (Nelson & Stenberg, 2017) -- we are unable to generalize to larger local governments in the United States or abroad.

This leads to a related issue of relative gender and racial homogeneity in the sample. Specifically. women and POC were underrepresented, especially in supervisor positions, indicating a glass ceiling effect towards the top of the organization (Krøtel, Ashworth, & Villadsen, 2019). Because women and POC were underrepresented, we must question how the effects identified in the present study would hold up in other more diverse contexts, or contexts where women and POC are the majority. To improve generalizability of future studies, replications could occur across occupations and sectors, at different levels of government, in regions of the world with different cultures, and in larger organizations. The extent to which a network is gendered is likely influenced by these different factors. Replicating this exploratory study across different environments would provide additional nuance to research on gendered tie formation. The cross-sectional nature of the data is another limitation, constraining our ability to infer causality between gender and discussion tie formation.

Future research should consider panel or survey experiments to provide a clearer depiction of gender's role in social network dynamics. Qualitative data could also be useful for identifying possible interim mechanisms driving tie formation: for example, discussion ties could be gendered if motivated by the desire to professionally network, or the quest to demonstrate one's knowledge, which could be construed as culturally masculine behaviors. Future survey instruments could ask questions about culture, leadership, and the nature of organizational roles. Information on the frequency of interactions would also enrich future analyses.

Finally, future social network analyses should offer an array of choices or open-ended responses for gender. Providing the binary choice of only male or female in the survey overlooks the increasing realities of public sector organizations as trans, non-binary, and gender-fluid people continue to join their ranks. Doing so may not be possible in every public organization given that cultural and political values vary by community but should be included whenever there is receptivity within the public organization to do so.

Additionally, this study lacks an intersectional perspective. Although initially used as a legal frame (Crenshaw, 1989), intersectionality is now a much broader framework used to explore interacting identities among organizational members within the context of public management scholarship (Bearfield, 2009; Hamidullah & Riccucci, 2017). While an intersectionality analysis is beyond the scope of this study due to the small number of women of color in the organization, future research that explores the intersection of gender and race would help refine discussions of tie formation, as the patterns we find with gender may be exacerbated when interacting with race. In addition, while public administration literature has primarily focused on race and gender when exploring intersectionality, scholars should also consider the wide range of identities present within public organizations. For instance, future research on network formation can explore immigration status, nationality, sexuality, or religious background.

Along with an intersectional perspective, future research should also consider the role of cultural competency within network analysis. Cultural differences have the potential to create additional barriers when attempting to establish network ties. Stemming from health care field and attempting to improve service to immigrant and racially diverse communities (Borrego & Johnson, 2012) cultural competence represents an organization or individual's ability to work in cross cultural settings (Cross et al., 1989). Within the context of work and employee relations, cultural competence can improve our understandings of network ties. For example, if one employee possesses a different cultural background amongst a group of organizational members with a shared cultural background, this could place that employee on the periphery of the organizational network and prevent them sending and receiving ties at a similar rate to their peers. Accounting for the various identities and cultures held by employees in public organizations is essential to understanding the complexity of network formation.

**CONCLUSION**

Gender dimensions of discussion tie formation are an important topic of future public administration research because the nature of tie formation has implications for organizational knowledge (Smith-Lovin & McPherson, 1993), information flow (Paruchuri & Awate, 2017), and career advancement (Cabrera & Thomas-Hunt, 2007). Given that public organizations at the local level are being called upon to solve complex and urgent problems facing their communities, addressing barriers that may prevent knowledge flow and organizational learning is central to serving the public effectively. Our exploratory findings of a social network analysis of single local government organization in the Southeastern U.S. suggest that gender, and the combination of gender and formal organizational status, may sometimes influence employee connections and network formation.

While the relationships detected in this study did not emerge as expected, they nonetheless raise interesting research questions for further consideration. Are men as willing as women to seek out others for discussion in other settings, in contradiction to expectations that they avoid the culturally feminine trait of vulnerability? Are women more desirable targets of discussion ties in other settings? If so, is this because cultural femininity construes women as more responsive, accommodating, and approachable than men, or some other reason? Do supervisors in traditionally hierarchical organizations receive fewer incoming discussion ties as a way for employees to dodge micromanagement? Are female supervisors really avoided as discussion ties because they violate norms of culturally feminine behavior? Future multi-method investigations into these questions will advance our understanding on the micro dynamics of intra-organizational networks, which in turn will inform our understanding of the information flows and subsequent knowledge creation needed for effective public sector organizations.

Hamidullah, M. F., & Riccucci, N. M. (2017). Intersectionality and family-friendly policies in the federal government: Perceptions of women of color. *Administration & Society*, *49*(1), 105–120.

Hamidullah, M. F., Riccucci, N. M., & Pandey, S. K. (2015). Women in city hall: Gender dimensions of managerial values. *American Review of Public Administration*, *45*(3), 247–262.

Heikensten, E., & Isaksson, S. (2019). Simon Says: Examining Gender Differences in Advice Seeking and Influence in the Lab. *Available at SSRN 3273186*.

Hochshild, A. R. (1983). The managed heart: Commercialization of human feeling. *Berkeley: The University of California Press. ANTA (2003). Shaping Our Future: Australia's National Strategy for Vocational Education and Training (VET). Brisbane: Australian National Training Authority*.

Ibarra, H. (1990). Differences in men and women's access to informal networks at work: An intergroup perspective. *National Meeting of the Academy of Management, San Francisco, CA*.

Ibarra, H. (1992). Homophily and differential returns: Sex differences in network structure and access in an advertising firm. *Administrative Science Quarterly*, 422–447.

Ibarra, H. (1993). Personal networks of women and minorities in management: A conceptual framework. *Academy of Management Review*, *18*(1), 56–87.

Ibarra, H. (1997). Paving an Alternative Route: Gender Differences in Managerial Networks. *Social Psychology Quarterly*, *60*(1), 91–102.
37

# Appendix

## Appendix 1. Survey Questions Used in the Study

### A1.1 Network Tie Formation Survey Question

Q30

How do you interact with the following department members in Administration? Please check all that apply. **If you do not interact with a person, please leave your response blank.**

|  | I seek out this person to discuss work-related topics. | I seek information from this person to do my job. | I seek permission from this person to do certain tasks. |
|---|---|---|---|
| Administration Employee 1 | ☐ | ☐ | ☐ |
| Administration Employee 2 | ☐ | ☐ | ☐ |
| Administration Employee 3 | ☐ | ☐ | ☐ |
| Administration Employee 4 | ☐ | ☐ | ☐ |
| Administration Employee 5 | ☐ | ☐ | ☐ |

### A1.2 Trust Questions

In thinking about trust in your workplace, how much do you agree or disagree with the following statements?

|  | Strongly disagree | Disagree | Somewhat disagree | Neither agree nor disagree | Somewhat agree | Agree | Strongly agree |
|---|---|---|---|---|---|---|---|
| In my department, employees trust supervisors. | ○ | ○ | ○ | ○ | ○ | ○ | ○ |
| In my department, supervisors trust their subordinates. | ○ | ○ | ○ | ○ | ○ | ○ | ○ |
| In my department, employees trust supervisors to make good decisions. | ○ | ○ | ○ | ○ | ○ | ○ | ○ |

**Appendix 2. Goodness of Fit Plots for ERGM**

The goodness-of-fit plots show a reasonably well-fitting model, given the included model parameters. The plots below are for Model 3.

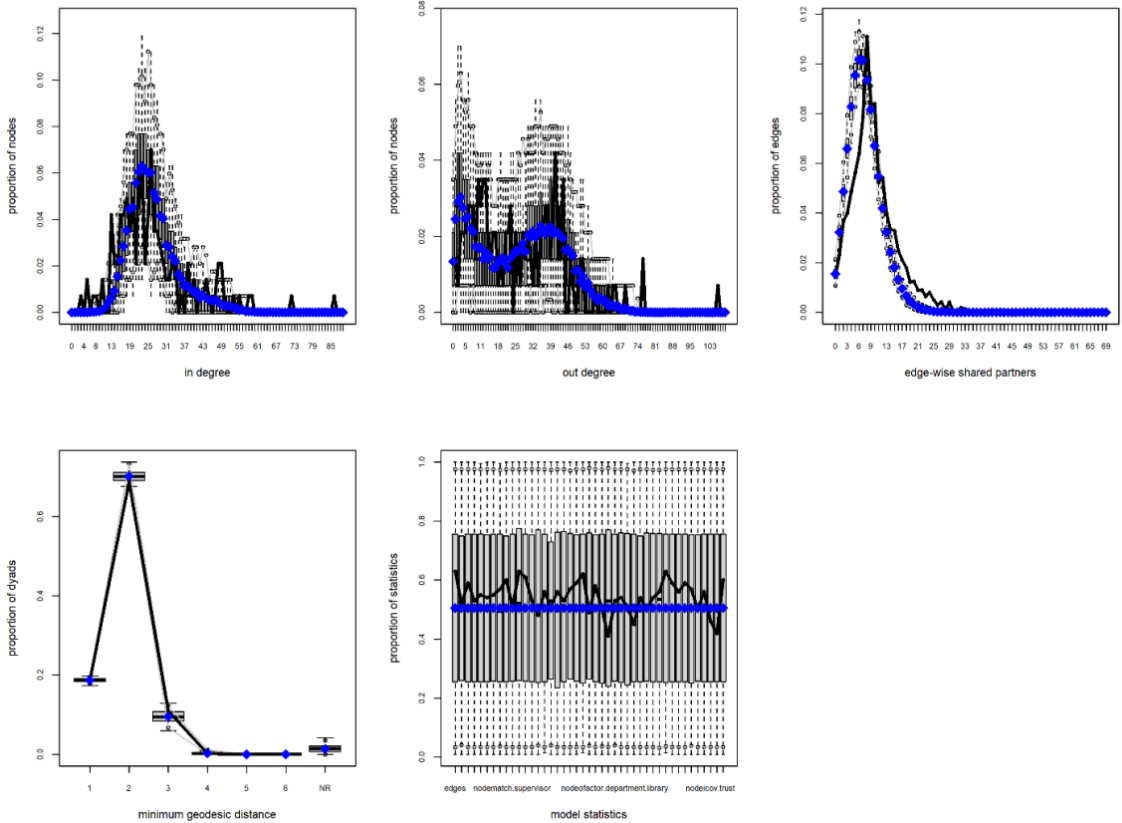